\documentclass[aps,epsfig]{revtex4}
\usepackage{graphicx}
\begin{document}
\def\be{\begin{equation}}
\def\ee{\end{equation}}
\def\bfi{\begin{figure}}
\def\efi{\end{figure}}
\def\bea{\begin{eqnarray}}
\def\eea{\end{eqnarray}}

\title{The fluctuation-dissipation relation in an Ising model without detailed balance}

\author{Natascia Andrenacci$^*$, Federico Corberi$^\dag$, and Eugenio Lippiello$^\ddag$}
\affiliation {Istituto Nazionale di Fisica della Materia, Unit\`a
di Salerno and Dipartimento di Fisica ``E.R.Caianiello'',
Universit\`a di Salerno,
84081 Baronissi (Salerno), Italy}

* andrenacci@sa.infn.it
\dag corberi@sa.infn.it
\ddag lippiello@sa.infn.it

\begin{abstract}

We consider the modified Ising model introduced by de Oliveira et al. [J.Phys.A {\bf 26}, 2317 (1993)],
where the temperature depends locally on the spin configuration and detailed balance 
and local equilibrium are not obeyed. We derive a relation between the linear response
function and correlation functions which generalizes the fluctuation-dissipation
theorem. In the stationary states of the model,
which are the counterparts of the Ising equilibrium states, 
the fluctuation-dissipation theorem
breaks down due to the lack of time reversal invariance. 
In the non-stationary 
phase ordering kinetics 
the parametric plot of the integrated response function $\chi (t,t_w)$ 
versus the autocorrelation function is
different from that of the kinetic Ising model. However,
splitting $\chi (t,t_w)$ into
a stationary and an aging term $\chi (t,t_w)=\chi _{st}(t-t_w)+\chi _{ag}(t,t_w)$,
we find $\chi _{ag}(t,t_w)\sim t_w^{-a_\chi }f(t/t_w)$, and
a numerical value of $a_\chi $ consistent
with $a_\chi =1/4$, as in the 
kinetic Ising model. 

\end{abstract}

\maketitle


\section{Introduction} \label{intro}

In equilibrium systems the integrated response function 
$\chi (t,t_w )$ and the autocorrelation function $C(t,t_w)$ 
depend only on the two time difference $\tau=t-t_w$ and are related 
by the fluctuation dissipation theorem (FDT)~\cite{nota}
\be
\chi (\tau )= \widehat \chi (C),
\label{fdtx}
\ee
where 
\be
\widehat \chi (C)=\beta [C(0)-C(\tau )]
\label{fdtorig}
\ee
and $\beta =-d\widehat \chi (C)/dC$ 
is the inverse equilibrium temperature.

In recent times many studies have considered the possibility to relate
response and correlation functions in non equilibrium
systems. In aging systems, such as glassy materials and coarsening systems, 
relaxation properties depend both on $t_w$ and $t$ and FDT breaks down.
In this context, guided by the solution of mean field spin glass models,
Cugliandolo and Kurchan proposed~\cite{CK} that, despite the explicit two-time dependence
of $\chi $ and $C$, a relation analogous to Eq.~(\ref{fdtx}), namely
\be
\chi (t,t_w)=\widehat \chi (C),
\ee
may still hold 
for a large class of systems in the large $t_w$ limit. The functional form
of $\widehat \chi(C)$, however, is different from the equilibrium one~(\ref{fdtorig}) and system dependent. 
In particular, it was shown~\cite{fmpp} that $\widehat \chi(C)$ is related
to basic properties of the equilibrium states; because of that, aging systems
can be classified~\cite{ruiz} into few classes according to the shape of $\widehat \chi(C)$.
Phase ordering systems are characterized by a broken line shape
of $\widehat \chi(C)$. More precisely, for $C$ larger than the
Edwards-Anderson order parameter
$q_{EA}$, FDT~(\ref{fdtorig}) still holds. For $C\le q_{EA}$ one has an horizontal line,
namely a constant integrated response function.  
In analogy to equilibrium systems, 
the quantity $-d\widehat \chi (C)/dC$ 
can be interpreted~\cite{peliti} as an {\it effective} inverse temperature $\beta _{eff} (C)$ of
the non equilibrium state. 
In coarsening systems this quantity takes 
two values, the temperature of the reservoir $\beta _{eff}(C)=\beta $, in the region $C>q_{EA}$, 
and $\beta _{eff}(C)=0$, for $C\le q_{EA}$. The feature $\beta _{eff}(C)=\beta$ in the region
of the largest values of $C(t,t_w)$ is quite general in systems where local 
equilibrium~\cite{kreuzer81} is obeyed. Local equilibrium, in fact, implies that on short timescales FDT holds.
Since small time separations 
corresponds to  the largest values of $C(t,t_w)$, in this regime one has $\beta _{eff}(C)=\beta $.
The situation is different in systems where local equilibrium is not obeyed. 
In this case one does not expect to observe FDT even in the short timescale regime
and the definition of a thermodynamic temperature from $\widehat \chi(C)$ may
be incorrect. 

In this paper we study the fluctuation dissipation relation in a two dimensional Ising model
without detailed balance (IWDB), originally introduced in Ref.~\cite{oliveira},
for which local equilibrium does not hold.
This spin system is analogous to the kinetic Ising model (KIM) but the temperature
entering the transition rates depends on space and time through the system
configuration. 
This model is known to behave much like the 
KIM. In the phase diagram, a disordered
high temperature phase and a low temperature phase with ergodicity breaking 
are separated by a critical line. The phase transition is characterized 
by the same critical exponents~\cite{oliveira} of the Ising model. 
After a quench into the low temperature phase, the non-stationary dynamics
is analogous~\cite{Godreche} to phase-ordering in the KIM. 

Despite these strong similarities,
lack of detailed balance makes the IWDB in principle
much different from the KIM and gives rise to new, interesting features that
can be enlightened by the fluctuation dissipation relation. In the KIM, stationary states
are equilibrium states characterized by time reversal invariance (TRI). 
Instead, in the stationary states of the IWDB, which are the counterparts of the KIM equilibrium states,
TRI is violated and FDT~(\ref{fdtorig}) breaks down. 
The relation between $\chi (t,t_w)$
and $C(t,t_w)$ may not be meaningful, as in systems with detailed balance,
and $-d\widehat \chi (C)/dC$ cannot be straightforwardly interpreted as a thermodynamic 
temperature. 
Interestingly, however, we can derive fluctuation-dissipation
relations that generalize to the present model what is known in systems 
where detailed balance holds.
In doing that, we uncover that the response function is not naturally related to the
spin autocorrelation function $C(t,t_w)=\langle \sigma _i(t)\sigma _i (t_w)\rangle$,
but rather to the quantity $A(t,t_w)=\langle \sigma _i(t)\sigma _i (t_w)\beta _i(t_w)\rangle$,
$\beta _i(t_w)$ being the space-time dependent inverse temperature.
In stationary states the fluctuation-dissipation relation reads  
\be
\chi (\tau )=A(0)-A(\tau )-\Delta (\tau ),
\label{fdtrr}
\ee
where $\Delta (\tau )$ is a quantity related to the lack of TRI, analogous to the
{\it asymmetry}~\cite{parisi} in systems with detailed balance out of equilibrium.
In the special case when the temperature is constant, $\beta _i (t)=\beta$,  
the model reduces to the KIM. Since now 
TRI is recovered, one has $\Delta (\tau )=0$ and then
\be
\chi (\tau )=A(0)-A(\tau ).
\label{fdt}
\ee
Recalling that in this case $A(\tau )=\beta C(\tau )$ one recognizes
the FDT, Eqs.~(\ref{fdtx},\ref{fdtorig}). In the case with non-constant temperature, 
since $A(t,t_w)\neq \beta C(t,t_w)$ both in stationary and non-stationary states,
the relation between $\chi (t,t_w)$ and $C(t,t_w)$ remains unclear.

In the phase-ordering process following a temperature quench, 
two time quantities can be splitted into
a stationary and an aging term. In particular, for the integrated response function
one has $\chi (t,t_w)=\chi _{st}(t-t_w)+\chi _{ag}(t,t_w)$,
where the aging part obeys the scaling form 
\be
\chi _{ag}(t,t_w)=t_w^{-a_\chi}f(y),
\ee
with $y=t/t_w$, as generally expected~\cite{Bouchaud}  in phase-ordering systems. 
When detailed balance holds, the exponent $a_\chi $ is
uniquely determined by the space dimensionality $d$ and the dynamic exponent $z$. 
Since in the IWDB the value of $z$ is the same~\cite{Godreche} of
the KIM, one expects the same value of $a_\chi $ in the two models. 
In fact, in $d=2$ we find results consistent with
$a_\chi =1/4$, as in the KIM.
This result complements those of Refs.~\cite{oliveira,Godreche} where it was
found that the {\it equilibrium} 
critical exponents and the {\it non-equilibrium} persistence exponent of the IWDB were the same, 
within statistical errors, of those of the 
Ising model. 
The numeric results of this paper, therefore, strengthens the
idea that the two models belong to the same non-equilibrium universality class. 

Despite this, the parametric plot of $\widehat \chi (C)$ 
is different from that of the KIM. Interestingly, the shape of this function is
similar to that found~\cite{Gonnella} in a 
soluble model of sheared
binary systems where detailed balance is also violated but local equilibrium
still holds. 
In particular, the extrapolation of the numerical data to the large
$t_w$ limit is consistent with the horizontal line typical of phase-ordering
in the region of small $C$ of the plot. 

This paper is organized as follows: In Sec.~\ref{sec2} we introduce the model. 
In Sec.~\ref{flucdiss} we derive a relation between the
response function and correlation functions which generalizes the FDT to the 
present model. This relation allows to compute numerically the response function
without applying a perturbation and to discuss the fluctuation-dissipation relation
in Sec.~\ref{num}. In particular, the stationary states at high and low temperature
are discussed in Secs.~\ref{eqsopra}, \ref{eqsotto}, while the aging dynamics following a quench
is considered in Sec.~\ref{neqsotto}. A summary and the conclusions are drawn in Sec.~\ref{concl}. 

\section{The model} \label{sec2}

We consider the Ising model defined by the Hamiltonian
\be
H[\sigma ]=-J\sum _{<ij>}\sigma _i \sigma _j=-\sum _i \sigma _i H_i [\sigma ],
\label{hamiltonian}
\ee
where $\sigma _i=\pm 1$ is a spin variable on a $d$-dimensional lattice and
$<ij>$ denotes nearest neighbors $i,j$ sites. $H_i[\sigma]=J\sum _{j_i}\sigma _{j_i}$,
where $j_i$ runs over the nearest neighbors of $i$, is the local field. 

A dynamics is introduced by randomly choosing a single spin on site $i$ 
and updating it in an elementary time step with a transition rate $w([\sigma]\to [\sigma'])$.
Here $[\sigma]$ and $[\sigma']$ are the spin configurations before and
after the move, which differ only by the value of $\sigma _i$.
In the IWDB transition rates are generic but their ratio must fulfill the condition
\be
\frac{w([\sigma]\to [\sigma'])}{w([\sigma ']\to [\sigma ])}=\exp \left \{-\sum _i \beta_i [\sigma ]
(\sigma _i H_i[\sigma ]-\sigma '_iH_i [\sigma '])\right \}.
\label{detbala}
\ee 
With a constant $\beta _i[\sigma ]=\beta $ one recovers the kinetic Ising model (KIM) in contact with a reservoir
at the temperature $T=\beta ^{-1}$. In this case Eq.~(\ref{detbala}) is the detailed balance condition
with respect to the Hamiltonian~(\ref{hamiltonian}). In fact, one has
\be
-\sum _i \beta (\sigma  _iH_i[\sigma ]-\sigma '_iH_i [\sigma '])=\beta (H[\sigma ]- H[\sigma '])
\label{sumham}
\ee
and hence 
\be
\frac{w([\sigma]\to [\sigma'])}{w([\sigma ']\to [\sigma ])}=\frac {\exp \left \{-\beta H[\sigma ']\right \}}
{\exp \left \{-\beta H[\sigma ]\right \}}=\frac {P_{eq}[\sigma ']}{P_{eq}[\sigma ]},
\label{detbal}
\ee 
where 
\be
P_{eq}[\sigma ]\propto \exp \left \{-\beta H[\sigma ]\right \}
\label{canonical}
\ee
is the canonical equilibrium probability.
Detailed balance implies that stationary states of the model 
are also equilibrium states with measure~(\ref{canonical}), which
are characterized by TRI. For a generic two-time quantity  $F(t,t_w)$, 
TRI implies $F(t,t_w)=F(-t,-t_w)$. If TTI is also obeyed, 
namely $F(t,t_w)=F(t-t_w)$, by shifting
time by an amount $t_w$, one also has $F(t,t_w)=F(\mid t-t_w \mid)$.  

In this paper we consider the case when $\beta _i[\sigma ]$ 
is not constant but depends on the configuration through the local field $H_i[\sigma ]$,
$\beta _i[\sigma ]=\beta (H_i [\sigma ])$. Physically, one can imagine a system in contact
with reservoirs at different temperatures each of which couples to the spins
$\sigma _i$ with the same local field $H_i[\sigma]$. 
Notice that flipping $\sigma _i$ does not change
$H_i[\sigma]$, hence $\beta _i[\sigma ]=\beta _i[\sigma ']$.
However, this is true only for the site $i$ where the flip occurs, while, in general,
 $\beta _j[\sigma ]\neq \beta _j[\sigma ']$ for $j\neq i$.
The very basic feature that makes this model different from the KIM
is the fact that its transition rates
do not obey detailed balance. This is expected on physical grounds, since
different local temperatures in the system determine heat fluxes that break TRI
and hence detailed balance.
Mathematically, this happens because the argument of the
exponential in Eq.~(\ref{detbala}) cannot be written as a difference 
${\cal H}[\sigma ']-{\cal H}[\sigma]$, ${\cal H}$ being a generic function,
as in Eq.~(\ref{sumham}), due to the factor $\beta _i[\sigma ]$. Indeed, the term
$\sum _i \beta_i [\sigma ] \sigma _i H_i [\sigma ]$ is a function ${\cal H}[\sigma ]$ 
of the configuration $[\sigma ]$ but
the term $\sum _i \beta_i [\sigma ]\sigma '_iH_i [\sigma ']\neq {\cal H}[\sigma ']$ because it 
depends on both the configurations $[\sigma ]$ and $[\sigma ']$.
Because detailed balance is not obeyed, 
the stationary states of the model~\cite{Godreche} are not equilibrium states and, in principle, TRI is
not expected. 

As discussed in~\cite{oliveira,Godreche,Chate}, the present model contains, 
as special cases corresponding to particular choices of
$\beta (\vert H_i \vert)$, the Voter, majority Voter and noisy Voter model, 
besides, clearly, the KIM.

\section{Fluctuation-dissipation relations} \label{flucdiss}

In this Section we derive a relation between the response function and particular 
correlation functions which generalize the result of Ref.~\cite{eugenio}
to the case of a non-constant $\beta _i[\sigma]$. The derivation closely
follows that of Ref.~\cite{eugenio} to which we refer for further details.

Let us consider a perturbing magnetic field
switched on the $j$-th site in  
the time interval $[t',t'+\Delta t]$, 
\be
h_i(t)=h \delta_{i,j}
\theta (t-t')\theta (t'+\Delta t -t)
\label{pert}
\ee
where $\theta $ is the Heavyside step function. 
The Hamiltonian~(\ref{hamiltonian}) is changed to 
\be
H[\sigma ]=-J\sum _{<ij>}\sigma _i \sigma _j-\sum _i h_i(t)\sigma _i=-\sum _i \sigma _i H^h_i [\sigma ],
\label{hamiltonianpert}
\ee 
where
$H^h_i[\sigma ]=J\sum _{j_i}\sigma _{j_i}+h_i(t)$.
In the limit of vanishing $h$, the effect of the perturbation~(\ref{pert}) on the spin on site $i$ at the time $t>t'$
is given by the linear response function~\cite{chat,crisanti}
\be
R_{i,j}(t,t')=\lim_{\Delta t \to 0} \frac{1}{\Delta t}
\left . \frac{\partial \langle \sigma _i(t) \rangle}{\partial h_j(t')} \right
\vert _{h=0},
\label{4}
\ee
where here and in the following $\langle \dots \rangle$ means ensemble averages, namely taken 
over different initial conditions and thermal histories.
Introducing the probability $p([\sigma ],t)$ to find the system in the configuration $[\sigma ]$
at time $t$, and the conditional probability 
$ p([\sigma],t\vert [\sigma '],t')$
to find the configuration $[\sigma ]$ at time $t$ given that the system was
in the configuration $[\sigma ']$ at $t'$, the r.h.s. of Eq.~(\ref{4}) can be written as 
\be
\left .\frac{\partial \langle \sigma_i(t) \rangle}{\partial h_j(t')} \right 
\vert _{h=0}=\sum _{[\sigma],[\sigma '],[\sigma '']}
\sigma _i p([\sigma],t\vert [\sigma '],t'+\Delta t) 
\left .\frac{\partial p^h([\sigma '],t'+\Delta t \vert [\sigma ''],t')}
{\partial h_j} \right \vert _{h=0} p([\sigma ''],t'). 
\label{5}
\ee
Here $p$ and $p^h$ refer to the conditional probabilities of the unperturbed and perturbed system,
respectively.
Let us concentrate on  the factor containing $p^h$.
The conditional probability for $\Delta t$ sufficiently small is given by
\be
    p^h([\sigma '],t'+\Delta t \vert [\sigma ''],t')= \delta_{[\sigma '],[\sigma '']}+
         w^h([\sigma '']\to [\sigma']) \Delta t + {\cal O}(\Delta t^2),
    \label{pippo}
\ee
where we have used the boundary condition $p([\sigma '],t\vert [\sigma ''],t)= 
\delta_{[\sigma '],[\sigma '']}$. 
Furthermore, also the perturbed transition rates $w^h$ 
must verify the condition~(\ref{detbala}), namely
\be
\frac{w^h([\sigma ']\to [\sigma''])}{w^h([\sigma '']\to [\sigma '])}=
\exp \left \{-\sum _i \beta_i [\sigma ']
(\sigma' _i H^h_i[\sigma ']-\sigma '' _iH^h_i [\sigma ''])\right \}.
\label{detbala2}
\ee  

Expanding the perturbed transition rates  in powers of $h$, one finds that
the following form 
\be
w^h([\sigma '] \to
[\sigma ''])=w([\sigma '] \to [\sigma ''])
\left  \{1-\frac{1}{2 }\beta_j [\sigma ']
(\sigma' _j h_j-\sigma ''_jh_j)\
\right  \}, 
\label{transh}
\ee
where $w([\sigma '] \to [\sigma ''])$ are generic unperturbed 
transition rates obeying~(\ref{detbala}), is compatible to first 
order in $h$ with the condition~(\ref{detbala2}).

Using Eqs.~(\ref{pippo}) and~(\ref{transh}), following Ref.~\cite{eugenio} 
the response function can
be written as the sum of two contributions
\be
R_{i,j}(t,t')=\lim _{\Delta t \to 0}\left [ D_{i,j}(t,t',\Delta t)+
 \overline{D}_{i,j}(t,t',\Delta t)\right ],
\label{7}
\ee
where 
\be
    D_{i,j}(t,t',\Delta t)=\frac{1}{2}
     \sum_{[\sigma],[\sigma ']}\sigma_ip([\sigma],t\vert
    [\sigma '],t'+\Delta t) \sum_{[\sigma '']\neq [\sigma ']}w([\sigma ']\to [\sigma ''])
\beta _j[\sigma '](\sigma '_j-\sigma ''_j)
    p([\sigma '],t')
\label{r1r1r1}
\ee
and
\be
\overline{D}_{i,j}(t,t',\Delta t)=\frac{1}{2}\sum_{[\sigma],[\sigma '],[\sigma '']\neq[\sigma ']}
\sigma_ip([\sigma],t \vert
[\sigma '],t'+\Delta t)(\sigma '_j-\sigma ''_j)\beta _j[\sigma '] w([\sigma '']\to [\sigma '])p([\sigma ''],t') .
\label{erre2}
\ee
Using the time translational invariance (TTI) of the conditional probability 
$p([\sigma],t\vert[\sigma '],t'+\Delta t)=p([\sigma],t-\Delta t\vert[\sigma '],t')$,
one can write $D_{i,j}(t,t',\Delta t)$ 
in the form of a correlation function
\be
D_{i,j}(t,t',\Delta t)=-\frac{1}{2} \langle \sigma_i(t-\Delta t)B_j(t')\rangle,
\label{cucu}
\ee
where
\be
B_j=-\sum _{[\sigma '']} (\sigma_j-\sigma ''_j)\beta _j[\sigma ] w([\sigma]\to [\sigma '']).
\label{bj}
\ee
Using Eq.~(\ref{pippo}), $\overline D_{i,j}(t,t',\Delta t)$ can be written as
\be
\overline{D}_{i,j}(t,t',\Delta t)=
\frac{1}{2}\frac{\Delta A_{i,j}(t,t')}{\Delta t}
\label{deltac}
\ee 
where
\be
 \Delta A_{i,j}(t,t')=\langle \beta _j(t')\sigma_i(t)[\sigma_j(t'+\Delta t)-\sigma_j(t')]\rangle
\label{re} 
\ee
Therefore, putting together Eqs.~(\ref{cucu}),(\ref{deltac}) and
taking the limit $\Delta t \to 0$ we obtain
\be
R_{i,j}(t,t')=\frac{1}{2}\frac {\partial A_{i,j}(t,t')}{\partial t'}-
\frac{1}{2}\langle \sigma_i(t)B_j(t')\rangle,
\label{nuova}
\ee
where 
\be
A_{i,j}(t,t')=\langle \beta _j(t')\sigma_i(t)\sigma_j(t')\rangle
\label{Aij}
\ee

In the following, we will be interested in the
integrated response function     
\be
\chi_{i,j}(t,t_w)=\int _{t_w}^t R_{i,j}(t,t')dt'
\ee
which correspond to the application of
a perturbing field between the times $t_w$ and $t$.
This quantity is easier to measure because switching on the perturbation for a finite time
increases the signal to noise ratio.
From Eq.~(\ref{nuova}) we have
\be
\chi_{i,j}(t,t_w) = \frac{1}{2}[A_{i,j}(t,t) - A_{i,j}(t,t_w)] -
\frac{1}{2}\int_{t_w}^{t}\langle \sigma_i(t) B_j(t') 
\rangle .
\label{new}
\ee
Eqs.~(\ref{nuova},\ref{new}) are the principal results of this Section.
They are relations between the 
response function and correlation functions of the unperturbed
kinetics, which generalize the FDT. These relations hold both in stationary and
non-stationary states, and do not depend on the choice of the unperturbed transition rates,
provided the condition~(\ref{detbala}) is obeyed. 

From Eq.~(\ref{new}) the integrated autoresponse function 
$\chi (t,t_w)=\chi_{i,i}(t,t_w)$, which does not depend on $i$ due to space
translation invariance, is given by
\be
\chi (t,t_w) = \frac{1}{2}[A(t,t) - A(t,t_w)] -
\frac{1}{2}\int_{t_w}^{t}\langle \sigma_i(t) B_i(t') 
\rangle ,
\label{new2}
\ee
where $A(t,t_w)=A_{i,i}(t,t_w)$. 
We will use Eq.~(\ref{new2}) for numerical
computations in the next sections. As discussed in~\cite{eugenio}
this method to compute $\chi (t,t_w)$ is much more efficient than traditional
methods where the perturbation is switched on. 

In stationary states, a simplified expression for $\chi (t,t_w)$ can be
obtained which makes the role of TRI evident. 
In order to do this let us consider the integral
\be
I(t,t_w)=\frac{1}{2}\int_{t_w}^{t}dt'\langle B_i(t) \sigma_i(t')\rangle 
\label{integral}
\ee
Enforcing Eq.~(\ref{bj}), proceeding as in~\cite{eugenio} the 
integrand can be written as
\be
\langle B_i(t)\sigma _i (t')\rangle=\frac {\partial \langle \beta _j (t)\sigma _j (t)\sigma _i (t')\rangle}
{\partial t}.
\label{ppp}
\ee
Using Eq.~(\ref{ppp}), replacing $d/dt$ with $-d/dt'$, due to TTI, and carrying out the integration
one has
\be
I(t,t_w)=\frac {1}{2}[A(t,t)-\langle \beta _i (t)\sigma _i (t)\sigma _i (t_w)\rangle ]
\ee
Adding and subtracting $I(t,t_w)$ on the r.h.s., Eq.~(\ref{new2}) 
can be cast in the form~(\ref{fdtrr}), with
\be
\Delta (\tau )=\frac{1}{2}\left \{
\langle \beta _i(t)\sigma _i(t)\sigma _i(t_w)\rangle-
\langle \beta _i(t_w)\sigma _i(t)\sigma _i(t_w)\rangle+
\int _{t_w}^t dt'\,\left [\langle \sigma_i(t)B_i(t')\rangle-
\langle B_i(t)\sigma_i(t')\rangle \right ]\right \}.
\ee

If TRI is also obeyed, as in equilibrium states, 
one has $\langle \beta _i(t)\sigma _i(t)\sigma _i(t_w)\rangle=
\langle \beta _i(t_w)\sigma _i(t)\sigma _i(t_w)\rangle$ and
$\langle \sigma_i(t)B_i(t')\rangle=\langle B_i(t)\sigma_i(t')\rangle$,
so that $\Delta (\tau )=0$.
Eq.~(\ref{fdtrr}) becomes a linear relation formally identical to Eq.~(\ref{fdt}). 
When TRI does not hold, instead, $\Delta (\tau )\neq 0$
and the relation between $\chi (\tau )$ and $A(\tau )$ is no longer linear.
As we will see in Sec.(\ref{stationary}), this is an efficient tool to check if
a stationary state is invariant under time reversal and, if not, to
quantify TRI violations. 

\section{Numerical results} \label{num}

In this section we present a numerical investigation of the dynamical properties of
the model and, in particular, of the fluctuation dissipation relation~(\ref{new2}).
We chose unperturbed transition rates of the Metropolis type for single spin flip on site $i$  
\be
w([\sigma ]\to [\sigma '])=\min \left \{1,\exp \{-\beta _i [\sigma] 
(H[\sigma ']-H[\sigma ])\}\right \},
\label{metropolis}
\ee
which, as can be easily checked, obey Eq.~(\ref{detbala}).
Up down symmetry implies that $\beta _i[\sigma ]$ does not depend on the 
sign of the Weiss field, $\beta _i[\sigma ]=\beta (\vert H_i \vert)$. 
In the following we will consider a system on a square lattice in two dimensions.
In this case the only possible values of
the local field are $H_i[\sigma ]/J=0,\pm 2,\pm 4$; the model
is then fully defined by assigning the three parameters $\beta(0),\beta(2J),\beta(4J)$.
Moreover, with the Hamiltonian~(\ref{hamiltonian}), 
the transition rates~(\ref{metropolis}) do not depend on $\beta (0)$. 
Then, at this level, the couple of inverse temperatures $\beta (2J),\beta (4J)$ is sufficient
to characterize the model. However, $\beta (0)$ becomes
relevant if the system is perturbed by an external magnetic field in order to
measure response functions.
Actually this quantity enters, through $A(t,t')$
and $B_i$, the expressions~(\ref{nuova},\ref{new}) of the response functions. 
Then, response functions
depend on $\beta (0)$, as already found numerically in Ref.~\cite{Chate}. 

\begin{figure}
\centering
   \rotatebox{0}{\resizebox{.5\textwidth}{!}{\includegraphics{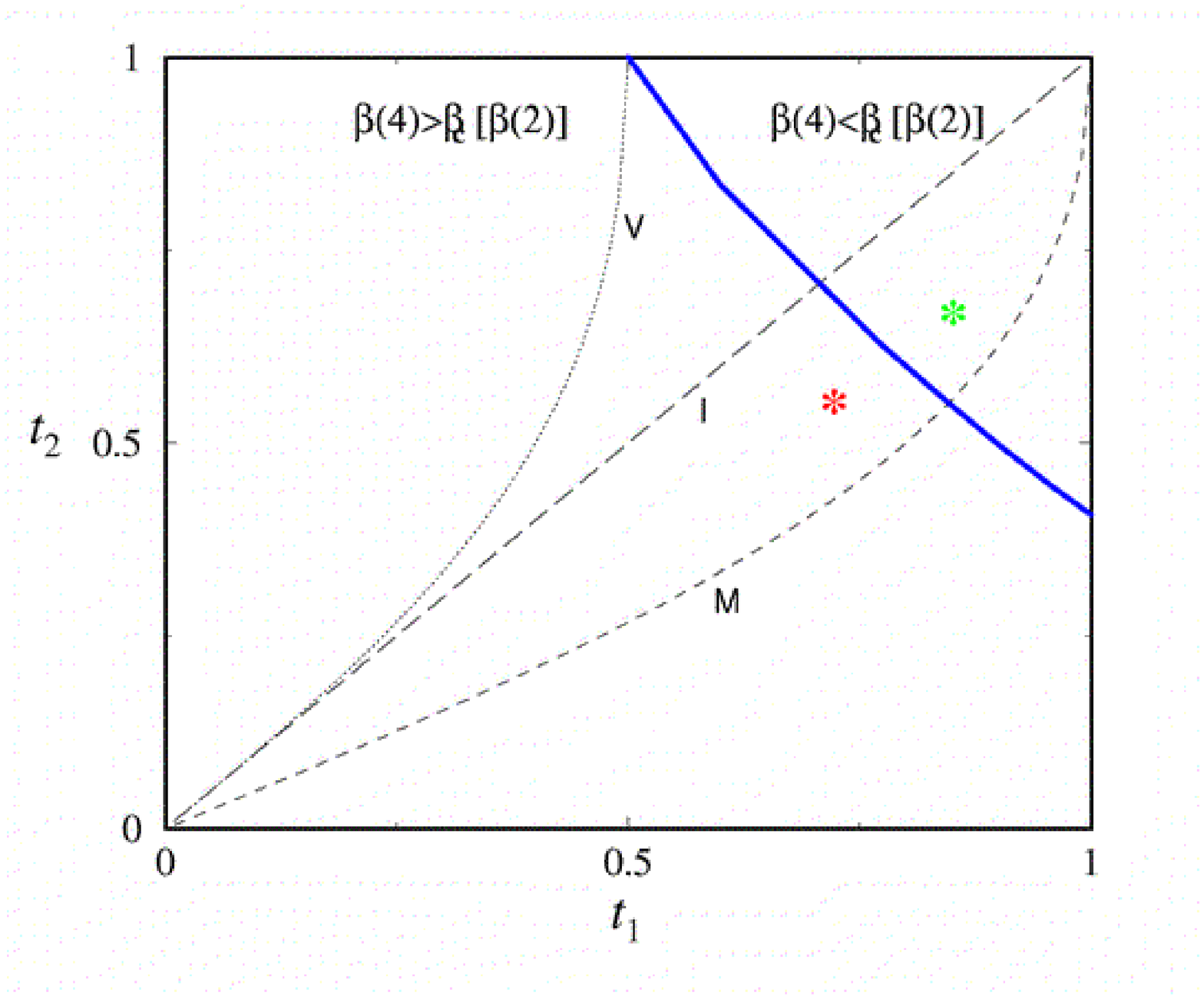}}}
\caption{The phase diagram of the IWDB model where $t_i=tanh(2 \beta_i)$. The three broken curves correspond respectively, V to the voter model, I to KIM and M to the extreme or  majority model. The two asterisks correspond to the two sets of parameters used in simulations. }
\label{figphasediag}
\vspace{2cm}
\end{figure}

The {\it phase diagram} of the IWDB was studied in Ref.~\cite{oliveira,Godreche}. It was shown
that in the plane of the parameters $\beta (2),\beta (4)$ one can identify two regions 
separated by a {\it critical } line $\beta (4)=\beta _c [\beta (2)]$, 
as shown in Fig.~(\ref{figphasediag}).
 The critical line starts at the inverse temperatures $\beta (2)=(1/2)$atanh$(1/2)$,
$\beta (4)=\infty$, corresponding to the voter model, where the transition
occurs in the absence of bulk noise, passes through the Onsager critical point with
$\beta (2)=\beta (4)= (1/2)$arcsinh$(1)$ and ends at $\beta (2)=\infty$, 
$\beta (4)\simeq 0.22$, corresponding to the extreme model,
where the transition occurs in the absence of interfacial noise.
For 
$\beta (4)<\beta _c [\beta (2)]$ one has an high temperature phase similar to the 
paramagnetic phase of the KIM.
Starting from any initial state the system quickly attains a stationary
state where the magnetization
\be
m=\langle \sigma _i \rangle ,
\ee
vanishes.  
For $\beta (4)>\beta _c [\beta (2)]$ there is a low temperature phase similar to
a ferromagnetic phase. Here there are two possible
dynamical situations, depending on whether the systems enters a state with broken symmetry,
namely with $m=\pm M_{BS}\neq 0$, or not. In the former case a stationary state 
is entered;
in the latter there is a phase-ordering process and the system ages.
We will consider these cases separately in Secs.~\ref{stationary},\ref{neqsotto},
where we will present the results of numerical simulations of a
two-dimensional system on a square lattice of size $1000 ^2$, with $J=1$.

\subsection {Stationary states}\label{stationary}

\subsubsection {$\beta (4)<\beta _c [\beta (2)]$} \label{eqsopra}

\begin{figure}
\centering
   \rotatebox{0}{\resizebox{.5\textwidth}{!}{\includegraphics{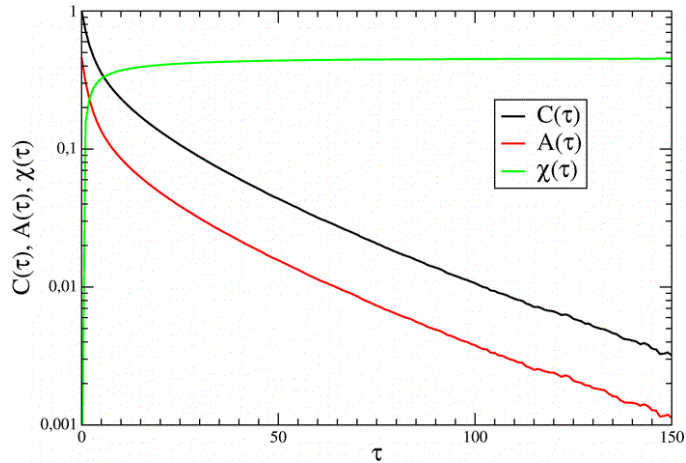}}}
\caption{$C(\tau ), A(\tau )$ and $\chi (\tau )$ are plotted against $\tau $.}
\label{figCAChieqsopra}
\vspace{2cm}
\end{figure}

We have prepared the system in the stationary state at the inverse temperatures
 $\beta (0)=0.80, \beta (2)=0.44, \beta (4)=0.30$, which correspond to the
paramagnetic phase. This state is quickly entered by the system by letting it
to evolve from any initial condition. 
In the stationary state we checked that $m=0$ and that  
two-time quantities
are functions of the time difference $\tau$
alone. In the following, time will be measured in montecarlo steps.
$C(\tau )$, $A(\tau )$ and $\chi (\tau )$ are shown in Fig.~\ref{figCAChieqsopra}. 

The behavior of $C(\tau )$ is analogous to that observed in the KIM. Starting
from $C(0)=\langle\sigma _i^2(t_w)\rangle=1$ the correlation function exponentially 
decays to zero. This is due to the decorrelation of the spin for large time differences 
$\lim _{\tau \to \infty}C(\tau )=\lim _{\tau \to \infty}\langle\sigma _i(t_w+\tau)\sigma _i(t_w)\rangle=
\langle \sigma \rangle \langle \sigma \rangle=m^2=0$.
Here we have introduced the simplified notation $\langle \sigma \rangle=\langle \sigma _i(t) \rangle$
to indicate that $\langle \sigma _i(t) \rangle$ does not depend on time $t$ nor on site $i$ due
to TTI and space homogeneity. We will use
this notation also in the following, dropping time and/or space variables whenever ensemble averages
do not depend on them.

$A(\tau )$ behaves similarly. From the definition (\ref{Aij}), its equal time value is the average inverse
temperature of the bath, $A(0)=\langle \beta \rangle=0.46$.
For large times difference also this correlation function decays to zero, since
$\lim _{\tau \to \infty}A(\tau )=m\langle \sigma  \beta \rangle=0$.  
Notice that $C(\tau )$ and $A(\tau )$ are proportional for large $\tau $ but not
for small $\tau $. This fact will be relevant in the following, when discussing
the possibility to define a thermodynamic temperature from the parametric plots
$\widehat \chi (C)$, $\widehat \chi (A)$ obtained plotting $\chi (\tau )$ versus 
$C(\tau )$ or $A(\tau )$, respectively. 

\begin{figure}
    \centering
   \rotatebox{0}{\resizebox{.5\textwidth}{!}{\includegraphics{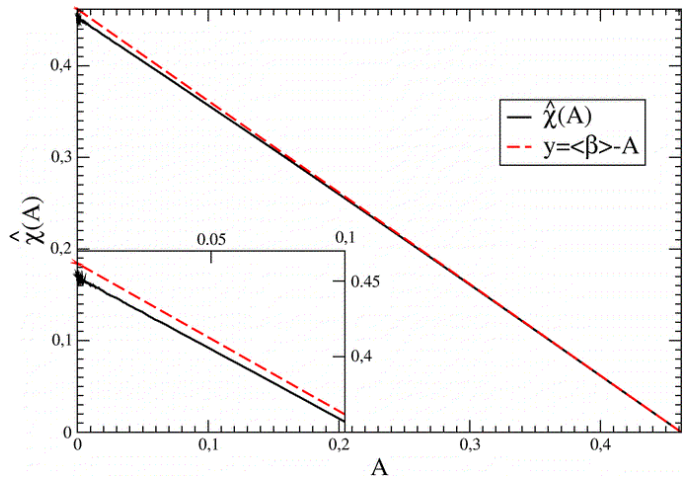}}}
    \caption{The parametric plot $\chi (A)$. In the inset the small $A$ sector is magnified.}
\label{figChidiAsopra}
\vspace{2cm}
\end{figure}

The behavior of $\chi (\tau )$ is also closely related to what is known for the KIM.
This quantity starts from $\chi (0)=0$ and saturates exponentially to a constant value
$\chi _\infty$. In the KIM this value is the equilibrium susceptibility, namely
the inverse temperature, $\chi _\infty=\beta $. One could conjecture that in the IWBD
this result can be generalized to $\chi _\infty=\langle \beta \rangle$.
However, as we will see shortly, this is not true due to lack of TRI.
In order to discuss this point let us consider,
in Fig.~\ref{figChidiAsopra} the parametric plot of $\widehat \chi (A)$.

For the largest values
of $A$, a relation formally identical to Eq.~(\ref{fdt}) is obeyed.
Recalling
Eq.~(\ref{fdtrr}) this implies that the term
$\Delta (\tau )$ is negligible. This, in turn, shows that TRI is satisfied in this
time domain.
This is reminiscent of what happens in out of equilibrium systems in contact with
a single reservoir, where the linear FDT relation~(\ref{fdt}) is found
on the r.h.s. of the $\widehat \chi (A)$ parametric plot, despite the system is not
in equilibrium. 
Since in this case $A(0)=\langle \beta \rangle$ Eq.~(\ref{fdt}) can be written as
\be
\widehat \chi (A)=\langle \beta \rangle -A. 
\label{avefdt}
\ee
This shows that the average bath temperature 
$\langle \beta \rangle$ enters the relation between $\chi (\tau )$ and $A(\tau )$,
as a natural generalization of what happens in the KIM where one has
$\chi (A)=\beta -A$.
This behavior can be explained
recalling that, in this sector of the plot, namely for $\tau \simeq 0$, the response
is provided by the fastest dynamical features. These are
the microscopic flipping of single spins that locally and 
instantaneously equilibrate at the current inverse temperature $\beta _i[\sigma ]$. 
Since the system is translational invariant, 
by taking ensemble averages one gets a sort of FDT
with respect to the average bath temperature $\langle \beta \rangle$, 
namely Eq.~(\ref{avefdt}). 
As larger values of $\tau $ are considered, corresponding to lower values of
$A(\tau )$, slower dynamical features are probed which cannot follow
the variations of $\beta _i[\sigma ]$.
In this sector, the contribution of 
$\Delta (\tau )$ becomes relevant and the parametric plot $\chi (A)$ deviates
from the straight line. Recalling the discussion of Sec.~(\ref{flucdiss}), 
$\Delta (\tau )\neq 0$ is related to the lack of TRI. 
These considerations, then, suggest that the parametric plot can be used as a convenient 
tool to detect and quantify TRI violations in non equilibrium stationary states.
Breakdown of TRI is also responsible for
the saturation of $\chi (\tau )$ to a value $\chi _\infty < \langle \beta \rangle$.  

\begin{figure}
    \centering
   \rotatebox{0}{\resizebox{.5\textwidth}{!}{\includegraphics{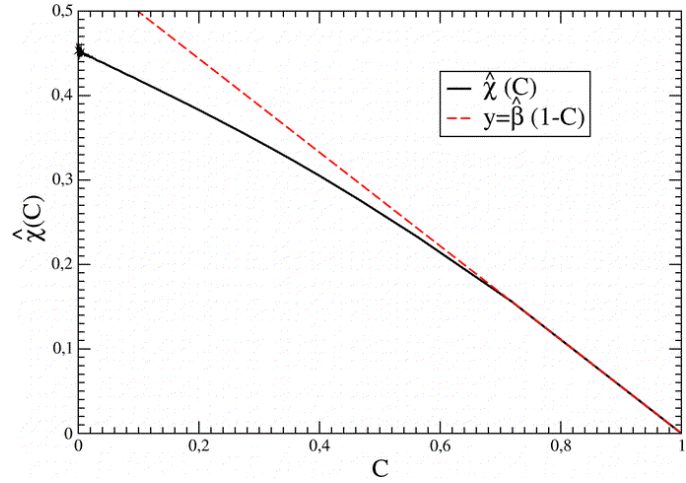}}}
    \caption{The parametric plot $\chi (C)$.}
\label{figChidiCsopra}
\vspace{2cm}
\end{figure}

The calculations of Sec.~\ref{flucdiss} clearly show that $A(t,t_w)$ is the
correlation function naturally associated to $\chi (t,t_w)$, rather then the autocorrelation
function $C(t,t_w)$ which does not enter the generalization of the
fluctuation dissipation theorem~(\ref{new2}). 
Nevertheless,we consider,
in Fig.~\ref{figChidiCsopra},
also the parametric plot of $\chi (\tau )$ versus $C(\tau )$, since, by analogy with
systems with detailed balance, this relation is often considered in the literature~\cite{Chate}.

Also in this case, for the larger values of $C$, the curve $\chi (C)$ obeys a linear
relation $\chi (C)=\hat \beta \cdot (1-C)$, with $\hat \beta =0.55$. 
Although this fact has suggested the interpretation~\cite{Chate} of $\hat \beta ^{-1}$
as a thermodynamic temperature, since $C(t,t_w)$ does not enter the fluctuation dissipation 
relation~(\ref{new2}), this reading is unmotivated. 
Notice in fact that $\hat \beta \neq \langle \beta \rangle$. Indeed, one should have
$\hat \beta=\langle \beta \rangle$ if $C(\tau)\propto A(\tau )$ in the regime considered,
namely for small $\tau $. Instead, as discussed previously, this is not the case.
Actually, we have checked that $\hat \beta \neq \langle \beta \rangle $
even when the choice of $\beta (0)$ proposed in~\cite{Chate} is
made, both in the high and low temperature phases.

For small values of $C$ the curve $\chi (C)$ strongly deviates from the straight
line. The parametric plot $\chi (C)$ can be compared to that found in stationary states
of other systems without detailed balance and, in particular, in an exactly soluble model of
binary systems under shear flow~\cite{Gonnella}, where a similar pattern was found.

\subsubsection {$\beta (4)>\beta _c [\beta (2)]$}\label{eqsotto}

\begin{figure}
    \centering
   \rotatebox{0}{\resizebox{.5\textwidth}{!}{\includegraphics{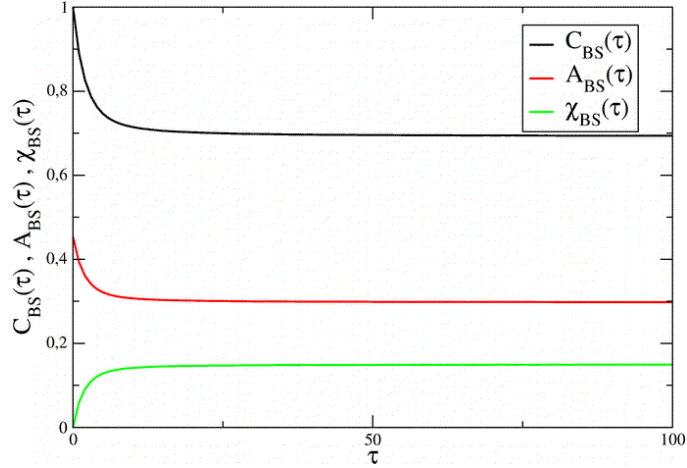}}}
    \caption{$C(\tau ), A(\tau )$ and $\chi (\tau )$ are plotted against $\tau $.}
\label{figCAChieqsotto}
\vspace{2cm}
\end{figure}

We have prepared the system in the stationary state at the inverse temperatures
$\beta (0)=0.80, \beta (2)=0.68, \beta (4)=0.37$, corresponding to the
ferromagnetic phase. This state is quickly entered by the system by letting it
to evolve from any initial condition with a broken symmetry $m\neq 0$. 
We consider the stationary state where the magnetization attains the positive value
$m=M_{BS}=0.83$.
Two time quantities, denoted by $C_{BS}(\tau), A_{BS}(\tau), \chi_{BS}(\tau)$
are plotted in Fig.~\ref{figCAChieqsotto}. 

\begin{figure}
    \centering
   \rotatebox{0}{\resizebox{.5\textwidth}{!}{\includegraphics{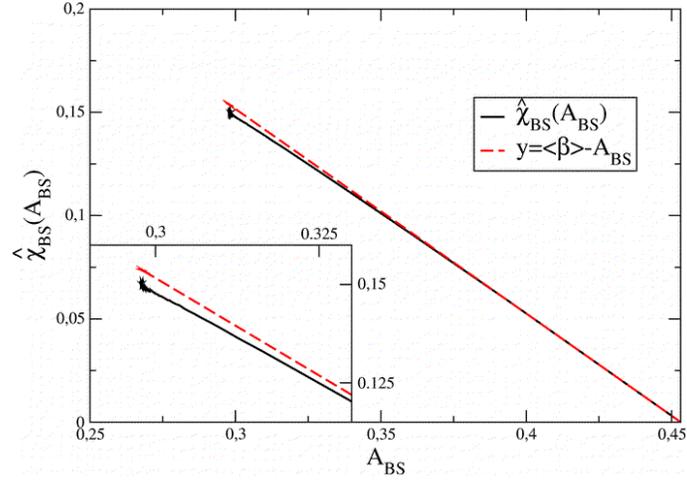}}}
    \caption{The parametric plot $\chi _{BS}(A_{BS})$. In the inset the small $A_{BS}$ sector is magnified.}
\label{figChidiAeqsotto}
\vspace{2cm}
\end{figure}

$C_{BS}(\tau )$ behaves similarly to the KIM. It decays
from $C_{BS}(0)=1$ to the large $\tau $ value
$C_{BS}(\infty )=
\langle \sigma \rangle \langle \sigma \rangle=M_{BS}^2=0.69$.
Analogously, $A_{BS}(\tau )$ decays from $A_{BS}(0)=\langle \beta \rangle=0.45$,
to $A_{BS}(\infty )=\langle \sigma \rangle \langle 
\sigma \beta \rangle=
M_{BS}\langle \sigma \beta \rangle=0.30$.
$\chi _{BS}(\tau )$ grows from zero up to the constant value
$\chi _\infty$. If TRI were obeyed, from Eq.~(\ref{fdtrr}) one should
have $\chi _{\infty }= \langle \beta \rangle-M_{BS}\langle \sigma \beta\rangle$.
Indeed, for the KIM this equation gives $\chi _{\infty }=\beta (1-M_{BS}^2)$
which is in fact the equilibrium susceptibility.
However, TRI is violated in the IWDB, as it is clear from 
the fact that the parametric plot of $\chi _{BS}(\tau )$ versus $A_{BS}(\tau )$, 
shown in Fig.~\ref{figChidiAeqsotto}, is not a straight line. 

Deviations from the straight line are due to 
the term $\Delta (\tau )$ in Eq.~(\ref{fdtrr}), which signals
the breakdown of TRI, and makes $\chi _{BS}(\tau )$ saturate to a value 
$\chi _\infty < \langle \beta \rangle-M_{BS}\langle \sigma \beta\rangle$.
Notice however that, also in this case, for the largest values
of $A_{BS}$, the linear relation $\chi _{BS}(A_{BS})=\langle \beta \rangle -A_{BS}$ is obeyed, as in
the paramagnetic phase, implying that TRI is satisfied in this
time domain and that the average temperature $\langle \beta \rangle$ can be extracted
from this region of the plot, in analogy with the KIM.

\begin{figure}
    \centering
   \rotatebox{0}{\resizebox{.5\textwidth}{!}{\includegraphics{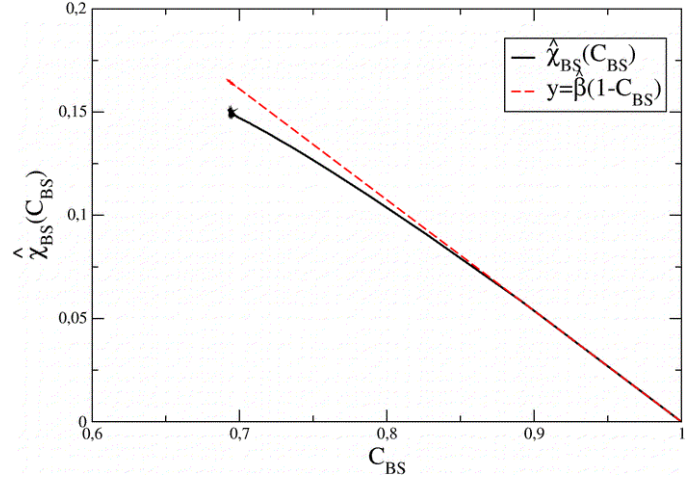}}}
    \caption{The parametric plot $\chi _{BS}(C_{BS})$.}
\label{figChidiCeqsotto}
\vspace{2cm}
\end{figure}

In Fig.~\ref{figChidiCeqsotto},
the parametric plot of $\chi _{BS}(\tau )$ versus $C_{BS}(\tau )$ is also shown.

This plot is analogous to the one found in the paramagnetic phase, and
similar considerations can be made. In particular,
we find
$\chi _{BS}(C_{BS})=\hat \beta \cdot (1-C_{BS})$, with $\hat \beta =0.54$, for the
largest values of $C$. As already discussed in Sec.~(\ref{eqsopra}), there is no
reason to interpret this quantity as an inverse temperature, 
and again $\hat \beta \neq \langle \beta \rangle$.

Let us also introduce the connected two time quantities, that will be used in 
Sec.~\ref{neqsotto}. Using the general definition of the connected correlation function ${\cal D}$ 
between two observables $O$ and $O'$, ${\cal D}=\langle OO' \rangle -\langle O\rangle
\langle O'\rangle$, the connected two-time quantities associated to $C_{BS}(\tau)$ and $A_{BS}(\tau)$ are
\be
{\cal C}_{BS}(\tau )=C_{BS}(\tau )-M_{BS}^2,
\ee 
\be
{\cal A}_{BS}(\tau)=A_{BS}(\tau )-M_{BS}\langle \sigma  \beta \rangle
\ee

\subsection {Aging dynamics}\label{neqsotto}

In this Section we study the non equilibrium process following a quench from an
initial disordered state with $m=0$ to the final inverse temperatures 
$\beta (0)=0.80, \beta (2)=0.68, \beta (4)=0.37$. Notice that these are the same
temperatures of the previous section~\ref{eqsotto}, corresponding to a point in the 
ordered phase. In this case one observes a phase ordering process where
domains of two phases with $m=\pm M_{BS}$ coarsen~\cite{Godreche}. In the interior of
such domains the system is found in the stationary state studied in the 
previous section. 
In analogy to what in known for the KIM, and more generally in aging systems~\cite{Bouchaud},
we expect quantities such as the equal time correlation function
\be
G(r,t)=\langle \sigma _i (t)\sigma_j (t) \rangle,
\label{gdierre}
\ee
$i$ and $j$ being two sites whose distance is $r$,
or two-time correlation functions and response 
to take the additive structure~\cite{Teff}
\be
G(r,t)=G_{st}(r)+G_{ag}(r,t),
\label{splitg}
\ee
\be
C(t,t_w)=C_{st}(\tau )+C_{ag}(t,t_w),
\label{splitc}
\ee
\be
A(t,t_w)=A_{st}(\tau )+A_{ag}(t,t_w),
\label{splita}
\ee
\be
\chi (t,t_w)=\chi_{st}(\tau )+\chi_{ag}(t,t_w).
\label{splitchi}
\ee
The presence of the stationary state in the bulk of the growing domains is the
origin of the contributions $G_{st}(r),C_{st}(\tau ),A_{st}(\tau ),\chi_{st}(\tau )$,  
while the terms $G_{ag}(r,t),C_{ag}(t,t_w),A_{ag}(t,t_w),\chi_{ag}(t,t_w)$
take into account the aging degrees of freedom in the system.

In the KIM quenched to the final temperature $\beta ^{-1}$, 
$G_{st}(r)$ is the correlation function
of the stationary state with broken symmetry at the same
temperature $\beta ^{-1}$, namely the correlation $G_{BS}(r)$.
We define $G_{st}(r )$ in complete analogy for the IWDB, $G_{BS}(r )$
being the quantity measured in the 
stationary state at the same inverse temperatures.
$G_{ag}(r,t)$ can then be obtained by subtraction, by using Eq.~(\ref{splitg}).
In the scaling regime $G_{ag}(r,t)$ obeys
\be
G_{ag}(r,t)=M^2_{BS}g (x),
\label{scalgr}
\ee
$x=\frac {r}{L(t)}$. 
This property will be tested below. 

\begin{figure}
    \centering
   \rotatebox{0}{\resizebox{.5\textwidth}{!}{\includegraphics{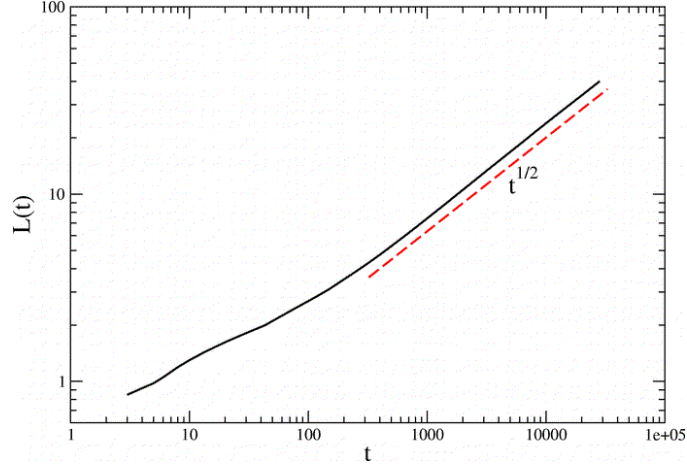}}}
    \caption{The typical size $L(t)$ of domains.}
\label{figlength}
\vspace{2cm}
\end{figure}

The typical size of domains can then be computed as the half height width of $G_{ag}(r,t)$.
This quantity is shown in Fig.~\ref{figlength}.
After the initial transient, $L(t)$ has a power law behavior $L(t)\sim t^{1/z}$.We measure $1/z=0.50$, as for the KIM.

\begin{figure}
    \centering
   \rotatebox{0}{\resizebox{.5\textwidth}{!}{\includegraphics{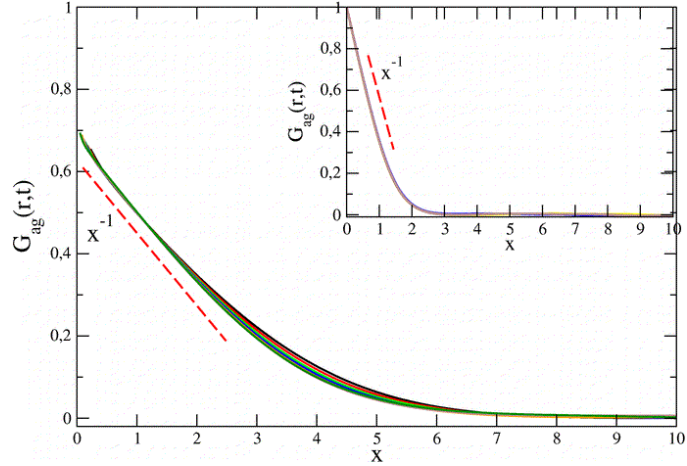}}}
    \caption{$G_{ag}(r,t)$ is plotted against $x$ for different times $t$ generated 
from $t_n=$Int$[\exp (n/2)]$ with 
$n$ ranging from 14 to 20. In the inset $G_{ag}(r,t)$ for the same values of $t$ is plotted for the KIM quenched at $T=0$. One observes the same small $x$ behavior $g(x) \sim 1/x$. }
\label{figscalG}
\vspace{2cm}
\end{figure}

Coming back to the scaling~(\ref{scalgr}), in order to check this form we plot in Fig.~\ref{figscalG} 
$G_{ag}(r,t)/M^2_{BS}$ against $x$ for different values of $t$. 
According to Eq.~(\ref{scalgr}) one should find data collapse for different times.
Actually, the collapse is good even if worst than in the KIM, particularly
for $x\simeq 0$. Notice also that the form of the scaling function $g(x)$ 
is very similar to that of the KIM.

Let us discuss now two-time quantities.
Analogously to what discussed above for $G(r,t)$,
in the KIM one has
\be
C_{st}(\tau )={\cal C}_{BS}(\tau ).
\label{stateq}
\ee 
We define $C_{st}(\tau )$ in complete analogy for the IWDB,
and the same is assumed for $A_{st}(\tau )$,
\be
A_{st}(\tau )={\cal A}_{BS}(\tau ).
\label{stateq2}
\ee 
For the integrated autoresponse function, in systems
with a constant $\beta $, $\chi _{st}(\tau )$ is the response produced in the bulk of domains 
and is defined as the quantity
related by Eq.~(\ref{fdtrr})
to the stationary parts of the correlation functions.
In this case $\Delta (\tau )=0$, Eq.~(\ref{fdtrr}) is the FDT~(\ref{fdt}) and  
\be
\chi _{st}(\tau) = 
A_{st}(0) - A_{st}(\tau)={\cal A}_{BS}(0) - {\cal A}_{BS}(\tau)
=A_{BS}(0) - A_{BS}(\tau).
\label{new4}
\ee 
Clearly, since $\chi _{st}(\tau) $ is related by FDT~(\ref{fdt}) to the correlation function of the 
broken symmetry equilibrium state, it is the integrated autoresponse function of that state.
Then one has
\be
\chi _{st}(\tau )=\chi _{BS}(\tau ).
\label{defchistat}
\ee
For the present model, in full analogy to the case with constant $\beta $, 
we use Eq.~(\ref{defchistat}) to define $\chi _{st}(\tau )$. Namely,
$\chi _{st}(\tau )$ is the quantity measured in the 
stationary state at the same inverse temperatures in the previous section. 
Let us now turn to discuss the properties of the aging contributions in 
Eqs.~(\ref{splitc},\ref{splita},\ref{splitchi}).
Recalling the behavior of $C_{BS}(\tau )$ one concludes that $C_{ag}(t,t_w)$ decays from 
$C_{ag}(t_w,t_w)=M_{BS}^2$ to zero, as in the KIM. In analogy to the KIM
we expect it to obey the scaling form 
\be
C_{ag}(t,t_w)=h_C(y),
\label{scalCag}
\ee
where $y=t/t_w$,
with the power law $h_C(y)\sim y^{-\lambda /z }$ for large $y$,
$\lambda $ being the Fisher-Huse exponent.
Analogously, given the behavior of $A_{BS}(\tau )$ discussed in the previous Section, 
one concludes that $A_{ag}(t,t_w)$ decays from
$A_{ag}(t_w,t_w)=M_{BS}\langle \sigma \beta \rangle=0.30$
to zero.  We expect a scaling form 
\be
A_{ag}(t,t_w)=h_A(y),
\label{scalAag}
\ee
as for $C_{ag}(t,t_w)$.
The response $\chi _{ag}(t,t_w)$ is produced by the interface degrees of freedom
whose number goes to zero during the ordering process.
For this reason, in the $d=2$ KIM this quantity after reaching a maximum
for $y \sim 1$ decays to zero. A scaling behavior is obeyed, namely  
\be
\chi _{ag}(t,t_w)=t_w^{-a_\chi}f(y),
\label{scalChiag}
\ee
with the power law $f(y)\sim y^{-a_\chi}$ for large $y$ and $a_\chi $ consistent with
$a_\chi =1/4$~\cite{noicosa}.
In order to check these scaling
forms we have extracted the aging terms as
\be
C_{ag}(t,t_w)=C(t,t_w)-C_{st}(\tau)=C(t,t_w)-{\cal C}_{BS}(\tau),
\ee
\be
A_{ag}(t,t_w)=A(t,t_w)-A_{st}(\tau)=A(t,t_w)-{\cal A}_{BS}(\tau),
\ee
and
\be
\chi_{ag}(t,t_w)=\chi(t,t_w)-\chi_{st}(\tau)=\chi(t,t_w)-\chi _{BS}(\tau).
\ee

\begin{figure}
    \centering
   \rotatebox{0}{\resizebox{.5\textwidth}{!}{\includegraphics{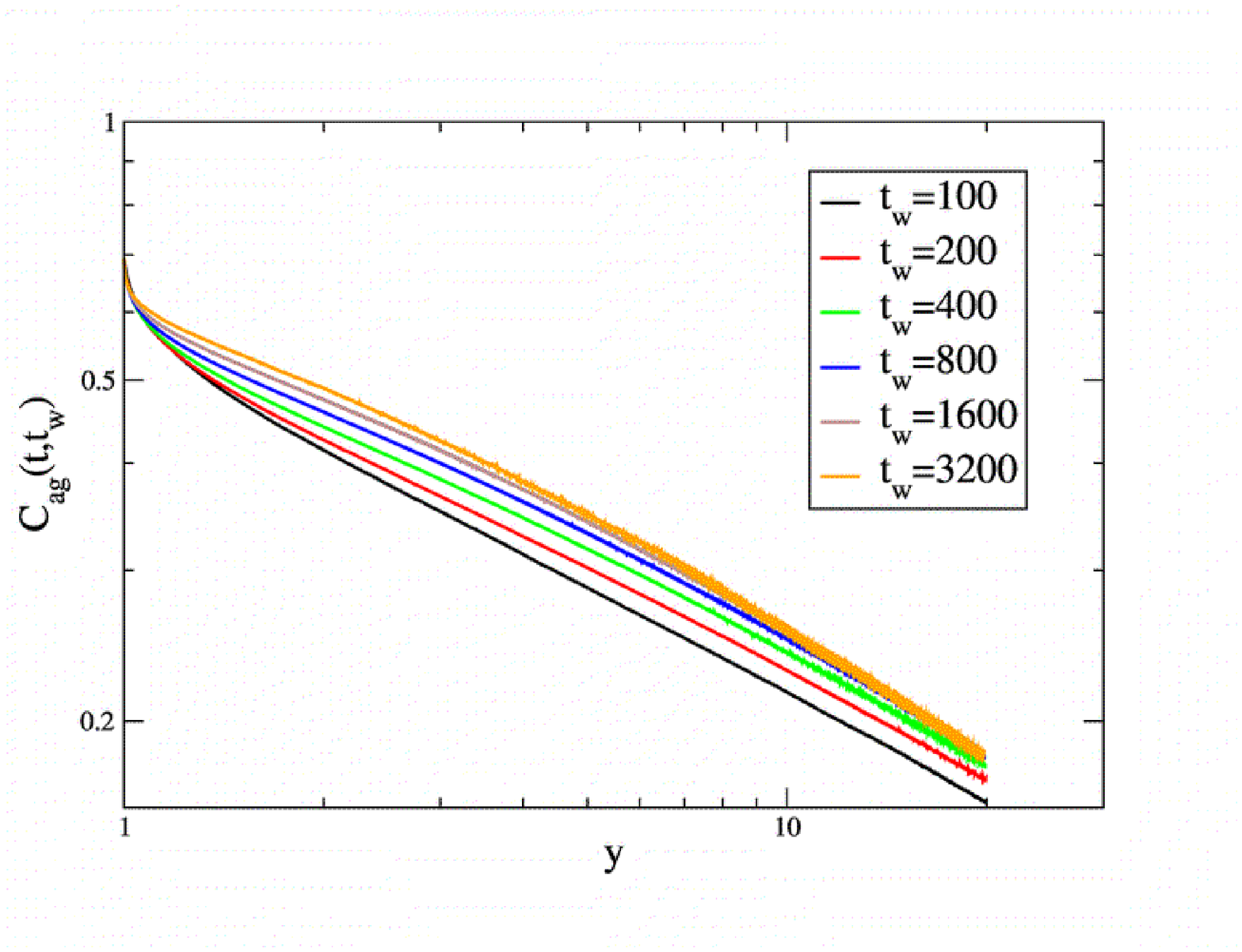}}}
    \caption{$C_{ag}(t,t_w)$ is plotted against $y$.}
\label{figCneqsotto}
\vspace{2cm}
\end{figure}

According to Eq.~(\ref{scalCag}), curves $C_{ag}(t,t_w)$ corresponding to different values of 
$t_w$ should collapse when plotted against $y$. This type of plot is shown
in Fig.~\ref{figCneqsotto}.

This figure shows that the data collapse is not very good for all times . This can be associated to
the presence of preasymptotic effects. However the quality of the collapse
gets better for the largest values of $t_w$ and $t$. Indeed, while the two curves with the
smallest values of $t_w$ ($t_w=100,200$) do not collapse at all, there is a tendency 
to a better collapse as $t_w$ gets larger. For the two largest values ($t_w=1600,3200$)
one has a nice collapse from $y\simeq 5$ onwards.

\begin{figure}
    \centering
   \rotatebox{0}{\resizebox{.5\textwidth}{!}{\includegraphics{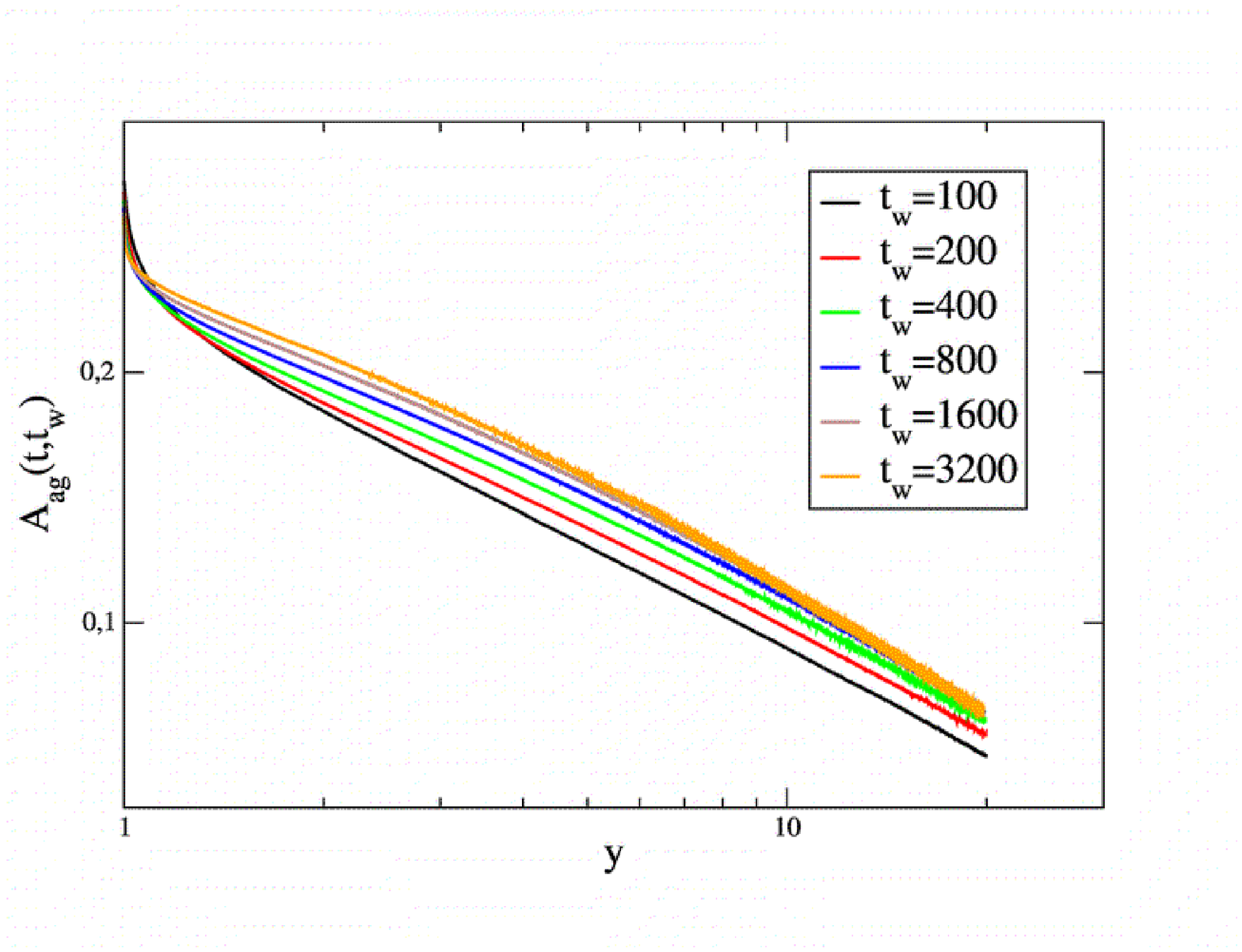}}}
    \caption{$A_{ag}(t,t_w)$ is plotted against $y=t/t_w$.}
\label{figAneqsotto}
\vspace{2cm}
\end{figure}

A similar situation is observed for the correlation $A_{ag}(t,t_w)$, as shown in 
Fig.~\ref{figAneqsotto}. Similarly to what found in stationary states, 
we find that $C_{ag}(t,t_w)\propto A_{ag}(t,t_w)$ for large $y$.

\begin{figure}
    \centering
   \rotatebox{0}{\resizebox{.5\textwidth}{!}{\includegraphics{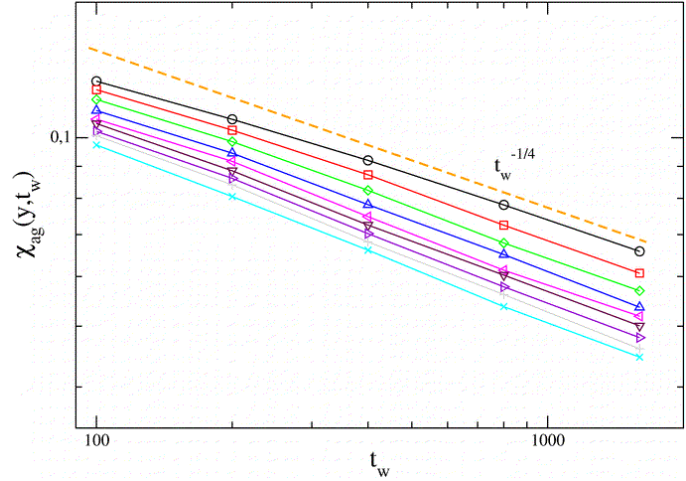}}}
    \caption{$\chi _{ag}(t,t_w)$ is plotted against $t_w$ for fixed $y=t/t_w$, and $y=3,5,7,9,11,13,15,17,19$ from top to bottom.  The dashed line is
    the power law $t_w^{-\frac{1}{4}}$.}
\label{figscalchisotto}
\vspace{2cm}
\end{figure}

\begin{figure}
    \centering
   \rotatebox{0}{\resizebox{.5\textwidth}{!}{\includegraphics{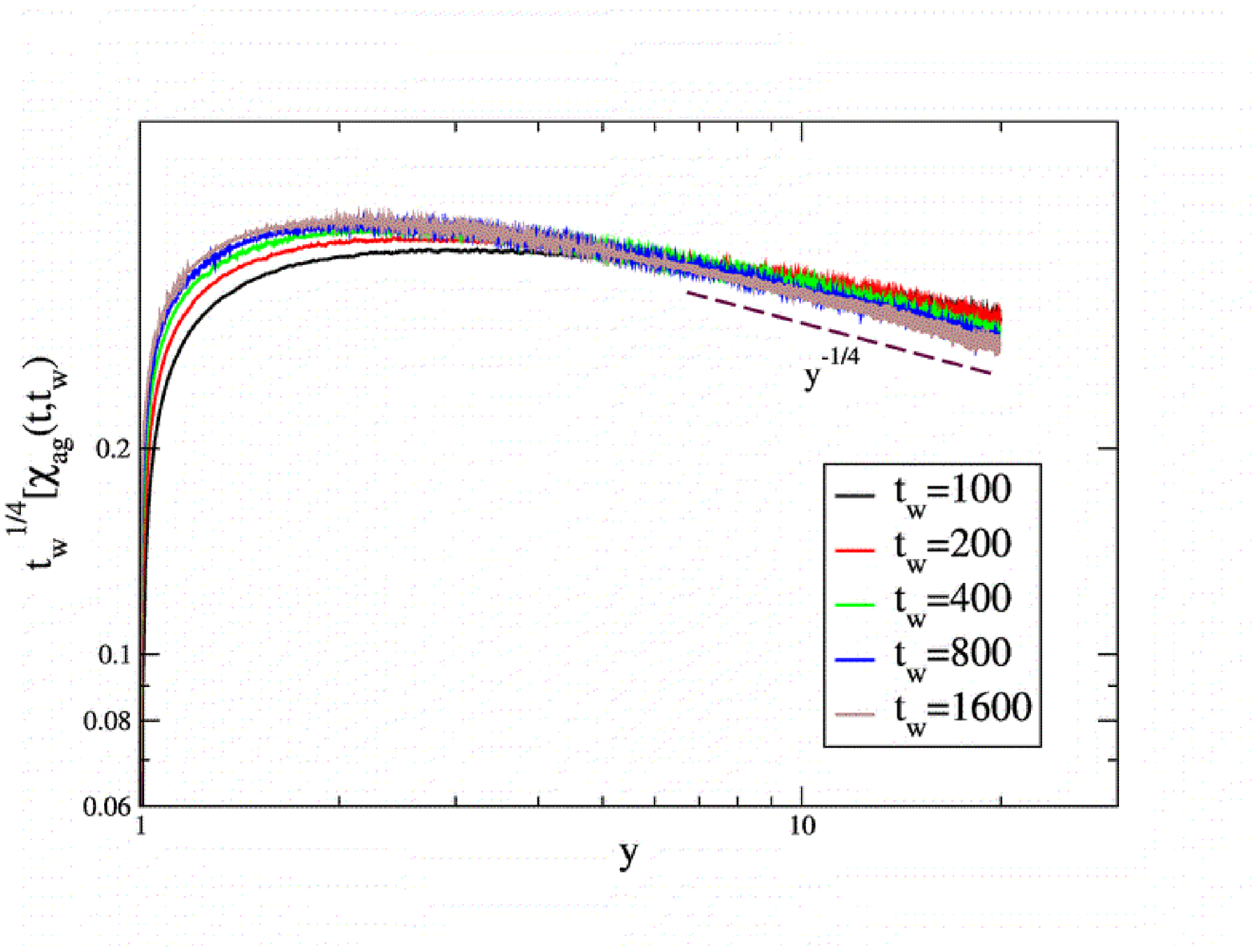}}}
    \caption{$t_w ^{\frac {1}{4}}\chi _{ag}(t,t_w)$ is plotted against $y=t/t_w$. The dashed line is
    the power law $y^{-\frac{1}{4}}$.}
\label{figChineqsotto}
\vspace{2cm}
\end{figure}

According to Eq.~(\ref{scalChiag}), the exponent $a_\chi $ can be extracted
as the slope of the double-logarithmic plot of $\chi _{ag}(t,t_w)$ against $t_w$, 
by keeping $y$ fixed.
We do this in Fig.~\ref{figscalchisotto}, for different choices of $y$.
We observe a good power law behavior, for every value of $y$. Best fit exponents 
are in the range $[0.23-0.28]$, depending on $y$, suggesting that the same value
$a_\chi =1/4$ of the KIM is found here.
Then, in order to check the data collapse,
in Fig.~\ref{figChineqsotto} we plot $t_w^{a_\chi }\chi _{ag}(t,t_w)$, with
$a_\chi =1/4$, against $y$, for different $t_w$.
The collapse is indeed rather good for the two largest values of $t_w$, implying that
Eq.~(\ref{scalChiag}) with an exponent consistent with $a_\chi =1/4$, as in the KIM, 
is asymptotically obeyed. 
This result complements those of Refs.\cite{oliveira,Godreche} where it was shown that
the IWDB has the same equilibrium critical exponent and the same persistence exponent
of the KIM. This strongly indicates that this two model belong to the same
{\it equilibrium} and {\it non-equilibrium} universality class.

\begin{figure}
    \centering
   \rotatebox{0}{\resizebox{.5\textwidth}{!}{\includegraphics{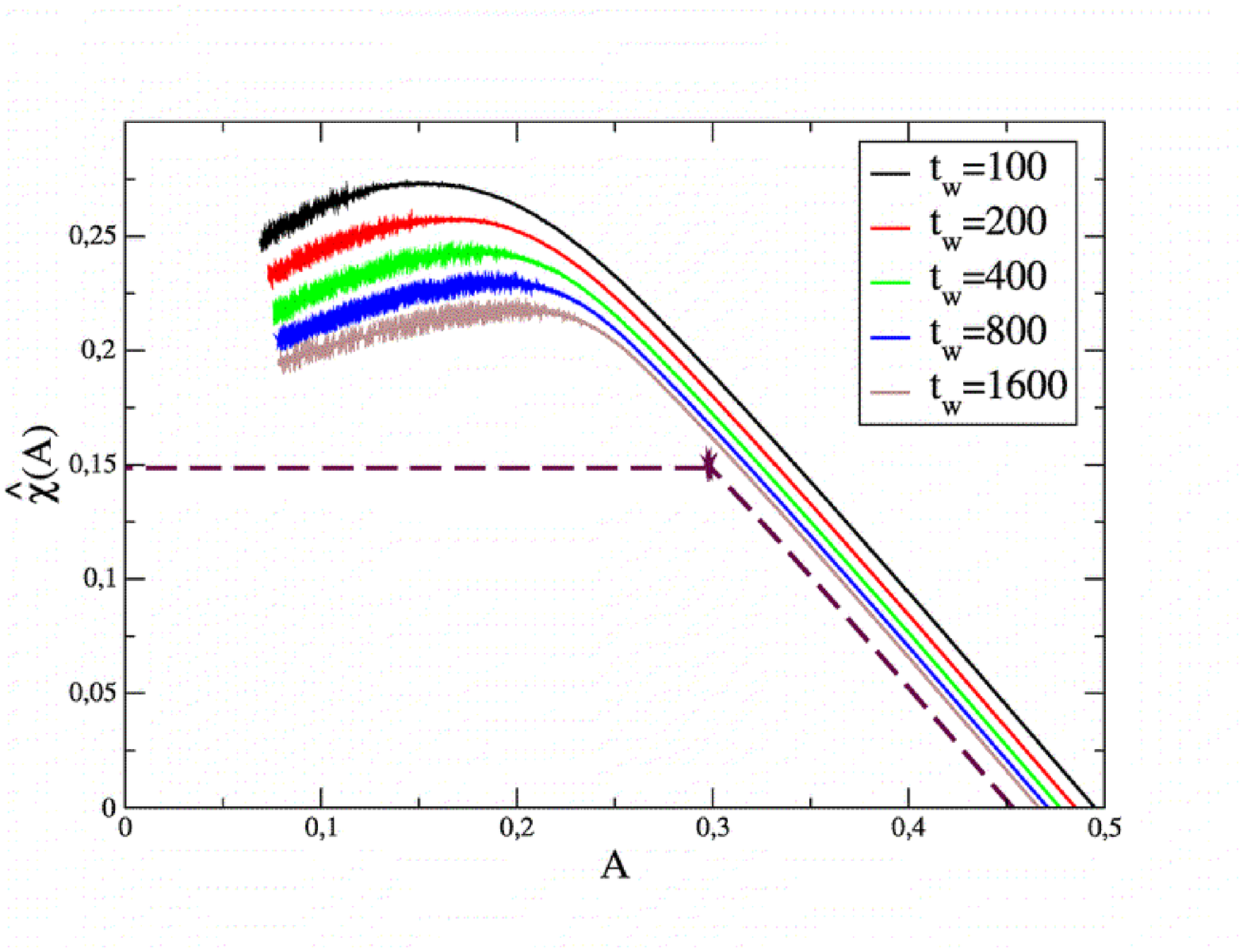}}}
    \caption{The parametric plot $\chi (A)$. The dashed line is the expected asymptotic
            behavior, for $t_w=\infty $.}
\label{figChidiAneq}
\vspace{2cm}
\end{figure}

In Fig.~\ref{figChidiAneq}, the parametric plot of $\widehat \chi (A)$ is shown.

In order to understand this plot one has to consider separately the
short time separation regime (ST), namely the limit $t_w \to \infty$ with $\tau /t_w \ll 1$,
and the large time separation regime (LT), where  $t_w\to \infty$ with $\tau /t_w \gg 1$.   
In the ST, given the scalings~(\ref{scalCag},\ref{scalAag},\ref{scalChiag}) 
the aging parts of two time functions remain equal to their
equal time value. For $A_{ag}(t,t_w)$ one has 
$A_{ag}(t,t_w)=A_{ag}(t_w,t_w)=
M_{BS}\langle \sigma \beta \rangle$.
Then, from Eq.~(\ref{splita}) one has $A(t,t_w)=
M_{BS}\langle \sigma \beta \rangle+A_{st}(\tau)=
M_{BS}\langle \sigma \beta \rangle+{\cal A}_{BS}(\tau)$. 
Recalling the behavior of
${\cal A}_{BS}(t,t_w)$ one concludes that, in the ST, $A(t,t_w)$ decays from $\langle \beta \rangle=0.45$
to $M_{BS}\langle \sigma \beta \rangle=0.3$.
In this time domain one has
$\chi _{ag}(t,t_w)=0$ and hence $\chi (t,t_w)=\chi_{st}(\tau)=\chi_{BS}(\tau)$.
Therefore, on the r.h.s. of the parametric plot of Fig.~\ref{figChidiAneq},
for $A\ge 0.3$, one should find exactly the same curve found in the stationary broken
symmetry state, namely Fig.~\ref{figChidiAeqsotto}. This curve is the broken line
in Fig.~\ref{figChidiAneq}. This implies that for the largest values of $A$, let's say for $A>0.4$,
Eqs.~(\ref{fdt},\ref{avefdt}) are obeyed, 
as discussed in Sec.~\ref{eqsotto}.

The numerical simulations can only 
access finite $t_w$, and some deviations from the asymptotic curve are then observed.
Notice, in particular, that the equal time value $A(t_w,t_w)$
has a weak time dependence. In this case there is a monotonous decrease of
this quantity. Recalling that $A(t_w,t_w)=\langle \beta (t_w)\rangle$
this means that in the kinetic process the average temperature is slightly
increasing. This happens because in the phase-ordering process the fraction
of bulk spins, with $\beta _i=\beta (4)$, grows in time. Since in this case 
$\beta (4)<\beta (2)<\beta (0)$ this corresponds to an increase of the average
temperature.
Despite these finite time effects however, the data clearly show that the curves 
converge to the broken line
increasing $t_w$. 

On the left hand side of the plot the LT is probed.
In this regime ${\cal A}_{BS}(t,t_w)=0$, and $A(t,t_w)=A_{ag}(t,t_w)$ decays from
$M_{BS}\langle \sigma \beta \rangle$ to zero. On the other hand $\chi_{st}(\tau)$
has already reached its asymptotic value $\chi_{st}(\tau)=\chi _\infty$ so that
$\chi (t,t_w)=\chi _\infty +\chi _{ag}(t,t_w)$. According to the scaling 
form~(\ref{scalChiag}), $\chi _{ag}(t,t_w)$ vanishes in the large $t_w$ limit.
Then, for $t_w=\infty$ on the left hand side of Fig.~\ref{figChidiAneq} one
should find the horizontal straight line typical of phase-ordering systems.
For finite values of $t_w$, $\chi _{ag}(t,t_w)$ still contributes to the
response and the curve overshoots the asymptotic value $\chi _\infty$.
However, as shown in Fig.~\ref{figChidiAneq}, the asymptotic
curve is approached increasing $t_w$. 

\begin{figure}
    \centering
   \rotatebox{0}{\resizebox{.5\textwidth}{!}{\includegraphics{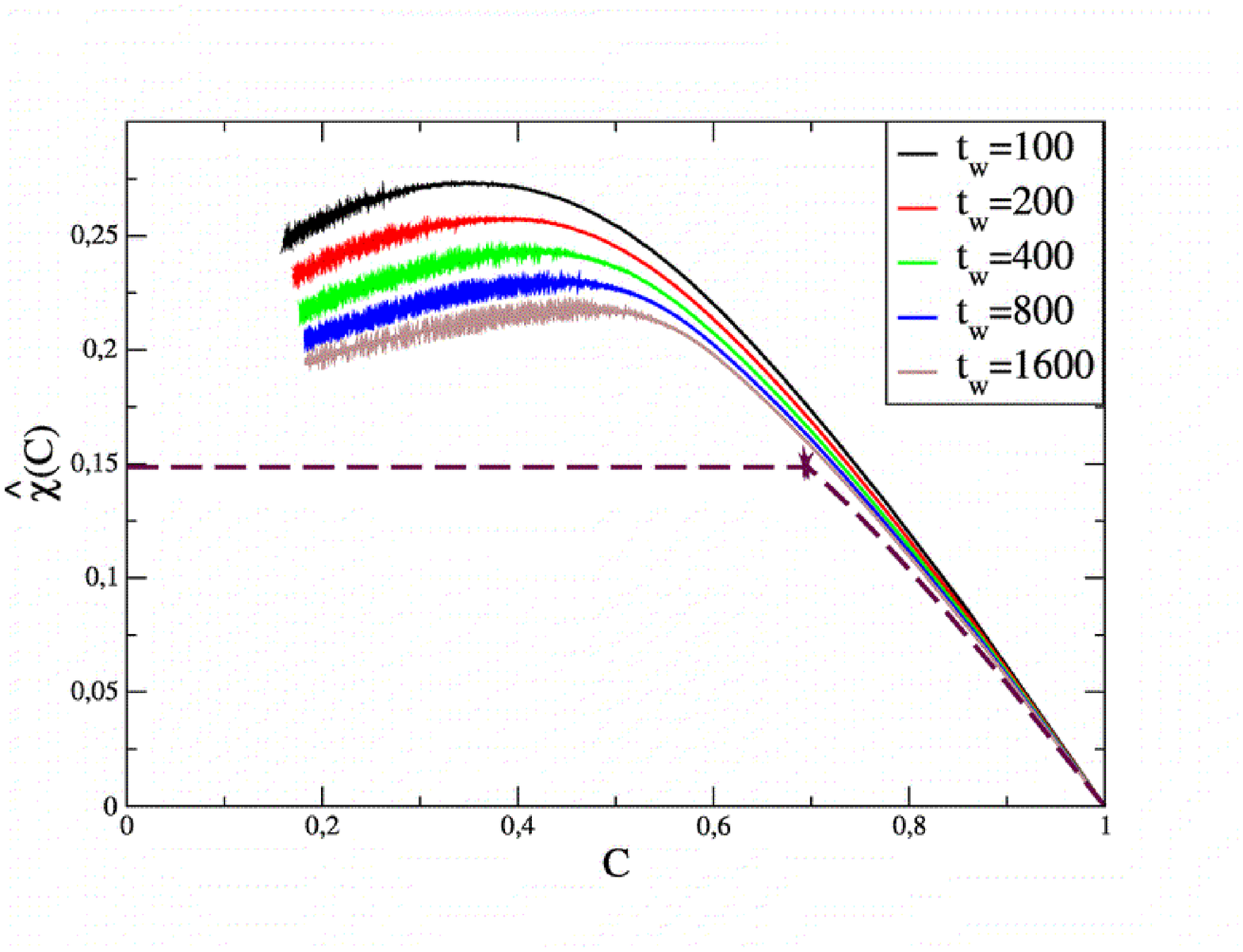}}}
    \caption{The parametric plot $\chi (C)$. The dashed line is the expected asymptotic
            behavior, for $t_w=\infty $.}
\label{figChidiCneq}
\vspace{2cm}
\end{figure}

Let us now consider the parametric plot of $\chi (t,t_w)$ versus $C(t,t_w)$,
shown in Fig.~\ref{figChidiCneq}. 

Repeating the same considerations as for the previous figure, one concludes that
on the r.h.s. of the figure the curves approach the curve of $\chi _{BS}(\tau)$
against $C_{BS}(\tau )$ of Sec.~\ref{eqsotto}, namely Fig.~\ref{figChidiCeqsotto},
in the large $t_w$ limit. We stress again that, as already discussed in Sec.~\ref{stationary}, 
there is no reason to associate the quantity $\hat \beta ^{-1}$ extracted from this sector
of the plot, to a thermodynamic temperature, as claimed in~\cite{Chate}.

On the l.h.s. of the figure our data are consistent with a convergence to the flat line 
$\chi (C)=\chi _\infty $, typical of phase-ordering.
Notice, however, that the whole shape of the parametric plot is different from
that of the KIM, due to the different relation between the stationary parts of $\chi (t,t_w)$
and $C(t,t_w)$, which shows up in the large-$C$ region.
Moreover, also in this case the
plot is similar to what observed in binary systems under shear~\cite{Gonnella}. 

\section {Summary and conclusions} \label{concl}

In this paper we have studied a modified Ising model, the IWDB, where the temperature entering the transition
rates depends on space and time through the system configuration and detailed balance 
is violated. This model is known to share many properties of the Ising model,
including the phase diagram, critical exponents~\cite{oliveira} and non-stationary dynamics~\cite{Godreche}. 

In systems with detailed balance a relation between the integrated response function $\chi (t,t_w)$, the
autocorrelation function $C(t,t_w)$
and the asymmetry $\Delta (t,t_w)$, a term related to the possible lack of
TRI, can be obtained under general assumptions~\cite{parisi,eugenio}.
This fluctuation-dissipation relation applies also in non-stationary states, namely out of equilibrium.

In this paper we have derived an analogous fluctuation-dissipation relation for the IWDB. 
The result is similar to the case with detailed balance but
the role played by $C(t,t_w)$ is now played by 
the correlation $A(t,t_w)$ between the spins
and the time dependent local inverse temperature. 
Since $\beta _i(t_w)$ enters the transition rates and is, therefore, correlated to the spin configuration,
$A(t,t_w)$ is not simply related to $C(t,t_w)$. 
It is therefore natural to consider the relation $\widehat \chi (A)$ for which a 
generalization of what is known in systems with detailed balance seems to
be possible, instead of the relation $\widehat \chi (C)$ whose meaning, as far as we can see,
remains unclear.   

In the stationary states of the model, which are the
counterparts of the Ising equilibrium states, the fluctuation-dissipation 
relation~(\ref{fdtrr}) is formally similar to the FDT~(\ref{fdt}), with the
important difference of a non vanishing $\Delta (t,t_w)$, determined by the violation of TRI.
This term makes $\widehat \chi (A)$ non linear. However, for
small time differences, namely for the largest values of $A$, the asymmetry can be
neglected and one recovers a linear relation, as in equilibrium systems,
with the average inverse temperature $\langle \beta \rangle$ playing the role of
$\beta $ in equilibrium systems. 
  
After quenching the systems into the ferromagnetic phase, a non stationary process
is observed, similar to the phase-ordering kinetics of the KIM.
We find that the response function exponent 
takes a value consistent with $a_\chi =1/4$, as for the KIM. This fact
complements previous results~\cite{oliveira,Godreche} on the universality between
the two models, both in {\it equilibrium} and {\it out of equilibrium}.
The shape of the  plots $\widehat \chi (A)$ and $\widehat \chi (C)$ 
can also be discussed in strict analogy to
what observed in the KIM. In particular, one finds the flat horizontal line
typical of phase ordering systems. Interestingly, the parametric plot $\widehat \chi (C)$
is similar to that of a soluble model of sheared
binary systems where detailed balance is also violated,
suggesting some qualitative similarity between these two model.

\vspace{1cm}
\centerline{\bf ACKNOWLEDGMENTS}
\vspace{1cm}
We warmly acknowledge Marco Zannetti for discussions and for the critical reading of the manuscript.

\vspace{.6cm}

This work has been partially supported
from MURST through PRIN-2004.


\begin{thebibliography}{000}
\bibitem{nota}
We cast a $\,\,\,\widehat {}\,\,\,$ over the response function to indicate
its dependence on the correlation function.
\end{thebibliography}
\end{document}